# High-*Q*/*V* air-mode photonic crystal cavities at microwave frequencies

Yinan Zhang[1], Irfan Bulu[1], Wai-Ming Tam[2], Ben Levitt[2], Jagdish Shah[2], Tancredi Botto[2] and Marko Loncar[1]

[1]*School of Engineering and Applied Sciences, Harvard University, Cambridge, USA*

[2]*Schlumberger-Doll Research Center, Cambridge, USA*

**Abstract**

We present results for a photonic microwave resonator designed and fabricated at 17.4GHz with a record high Quality factor ($Q$ = 26,400) at room temperature over a mode volume smaller than one cubic wavelength. The cavity is uniquely designed to have its electric field concentrated in air, which allows for efficient coupling to free space and facilitates interactions with gaseous atomic systems and fluids.

**1. Introduction**

Photonic crystal cavities with high Quality factors ($Q$s) are capable of extending the lifetime of cavity photons confined within mode volumes ($V$) of the order of a cubic-wavelength, and hence can greatly enhance the interaction between electromagnetic fields and matter. Recently, high $Q/V$ photonic crystal cavities have been extensively investigated with various geometries at optical frequencies, acting as a powerful platform for studying cavity quantum electrodynamics (cQED) [1] and developing a myriad of nano-photonic devices. Yet few photonic crystal cavities have been reported at microwave frequencies [2-7].

A high $Q/V$ microwave cavity is demanded for a number of applications. In atomic physics, a single Rydberg atom coupled to a single photon is the fundamental system for cavity QED studies, especially when one reaches the interesting strong coupling regime where energy can be coherently exchanged between the two. For strong coupling to occur, the atom-photon coupling rate $g$ (inversely proportional to square root of $V$) must exceed both the photon leakage rate (inversely proportional to $Q$) and the atom decoherence rate. Since the frequency difference of Rydberg states is in the microwave region [8], a high $Q/V$ microwave cavity is crucial to achieve this type of coupling. Strong coupling in Rydberg atoms has been previously observed in large-scale Fabry-Perot cavities and superconducting circuits [8, 9]. Photonic crystal cavities, in contrast, offer an alternative platform that is more compact and can be operated at room temperature.

Secondly, given the cavity's intrinsic field enhancement factor, there are also significant advantages in practical applications such as tunable microwave filters and antennas, where the efficiency of a radiator can be significantly improved by coupling with the cavity mode [10]; fluorescence microscopy for biological systems, where a high $Q/V$ microwave cavity can be used to control the temperature and drive the biological reactions [11]; resonance-enhanced microwave detectors [3]; particle accelerators [5, 6]; and refractive index sensors.

In this work, we have designed and experimentally demonstrated an all-dielectric photonic crystal microwave cavity. The cavity can be fabricated with a variety of materials using conventional machining techniques. We report a cavity mode at 17.4GHz with a very small mode volume limited to about one cubic-wavelength and a record-high $Q$-factor of 26,400 at room temperature. Besides the high $Q/V$ factor, the defect mode has a TM-polarized electric field concentrated in the air region [12], as opposed to most photonic crystal cavities designs with TE-polarized electric field concentrated in the host dielectric [13]. These features are crucial for effectively coupling microwaves to matter in a number of applications. Moreover, when a center-fed antenna is placed inside the cavity acting as a radiating dipole, we observe a strong signature of the cavity's Purcell enhancement factor. In summary, we believe this type of device offers great promise for studying cavity QED phenomena as well as for enabling novel applications at microwave frequencies.

## 2. Cavity design

The structure we propose starts from a so-called rectangular dielectric rod waveguide with a periodic array of circular holes, as shown in Fig. 1(a) [13-17]. We choose a refractive index of 3.1 for the rod, which corresponds to the refractive index of alumina ($Al_2O_3$) at microwave frequencies. The rectangular waveguide has an aspect ratio (thickness/width) of 2:1, and supports a fundamental TM-polarized mode with its major component aligned along the $y$-axis [Fig. 1(c)]. From our previous work [13], introducing a suitable periodic array of holes in the bulk materials of this high aspect-ratio waveguide results in a relatively wide bandgap for TM-polarized modes. Here we select the thickness to be $w$ = 4.72mm, depth $h$ = 9.43mm, and hole periodicity $a$ = 4.15mm. The radius of the holes is chosen as $r$ = 1.25mm.

Next, we introduce a defect region into the cavity by gradually increasing the periodicity (hole-to-hole distance) and hole diameter for each segment starting from a pair of outer holes and symmetrically moving towards the center. When the feature size of a segment is enlarged, the band-gap is red-shifted, resulting in a graded photonic band, as shown in Fig. 1(b). This allows confining an air-band mode in the defect region: the air-band mode is coupled to the evanescent Bloch modes within the band-gaps at each end, effectively trapping it between a pair of Bragg mirrors.

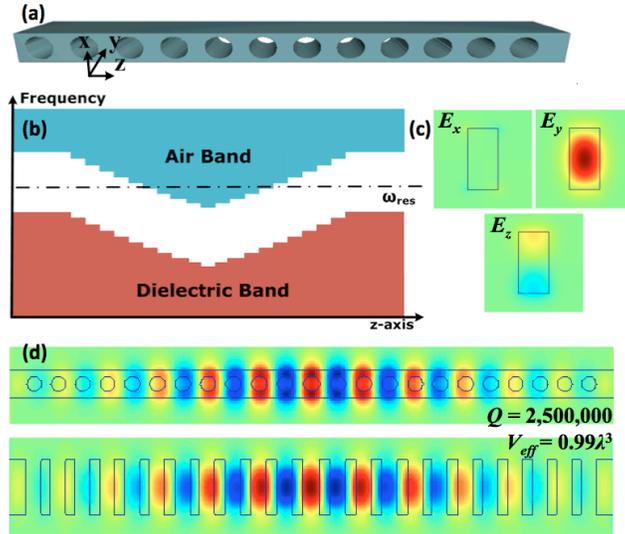

Fig. 1. (a) Schematic of the device. (b) Diagram of tapered photonic bandgap. (c) Transverse mode profile of the fundamental TM-polarized mode. (d) Mode profile of the cavity mode.

A three-dimensional finite-difference time-domain (FDTD) code was used to calculate the resonant frequency and the $Q$ factor of the cavity mode. On each side of the cavity we introduce 10 segments with progressive tapering of the hole diameter and spacing, in order to adiabatically couple the cavity mode to the evanescent Bloch mode. This adiabatic process is tuned to suppress the scattering loss resulting from the effective index mismatch between the two modes [18, 19]. Without material losses, the cavity mode has an ultra-high $Q$ factor of 2,500,000, limited by scattering alone ($Q = Q_{sc}$). The cavity mode profiles are shown in Fig. 1(d), where the $E_y$ component is plotted at the mirror plane of $y$-axis and $x$-axis, respectively. The tapered section's parameters are tailored to localize an air-band mode resonating at 17.4GHz by maximizing confinement along the z-axis, which results in a small cavity mode volume. As expected from the general features of air-band modes, the electric field concentrates in the air-region. However, the maximum of the electric field density [$\varepsilon|E|^2$] is not in the air region, so we define the effective cavity mode volume as,

$$V_{eff} = \frac{\iiint \varepsilon |E|^2 dV}{|E_{max,air}|^2} \quad (1)$$

where $E_{max,air}$ is the electric field maximum in the air region, located at the center of the cavity. For our cavity we found an effective mode volume of $0.99\lambda^3$.

The cavity mode is normally excited from an external waveguide port. The amount of energy stored ($U$) is proportional to the launched power ($P_0$), according to [20],

$$U = \frac{2}{\omega_0} \frac{Q^2}{Q_w} P_0 \quad (2)$$

In Eq. 2 $Q_w$ denotes the quality factor responsible for the field energy leakage from the cavity section via the dielectric waveguide. The total cavity quality factor $Q$, including material losses, can be written as

$$\frac{1}{Q} = \frac{1}{Q_{sc}} + \frac{1}{Q_m} + \frac{1}{Q_w} \quad (3)$$

where $Q_{sc}$ accounts for scattering losses due to mode mismatch in the tapered section and $Q_m$ for material losses in the dielectric. Substituting Eq. 2 into Eq. 1, we have

$$U = U_{max} = \frac{P_0}{w_0} Q, \text{ when } Q_w = \frac{Q_{sc} Q_m}{Q_{sc} + Q_m} \quad (4)$$

Therefore, for given $Q_{sc}$ and $Q_m$, it is important to tune $Q_w$ by fixing the number of Bragg mirror pairs on each side of the resonator so as to achieve optimal critical coupling conditions by satisfying Eq. 4.

## 3. Fabrication and measurement

The resonator was fabricated with ultra-high purity alumina from Coorstek (AD998 PlasmaPure). This material has a relatively high refractive index of 3.1 at microwave frequencies and very small dielectric losses. Dielectric losses are important as they ultimately limit the quality factor of the resonator. The material that we use has a nominal loss factor of $\tan(\delta) = 2.5 \times 10^{-5}$ at 6GHz, which results in $Q_m = 44,000$. For our device, we put 14 Bragg mirror pairs at each side, which corresponds to $Q_w = 210,000$. Therefore, the total $Q$ factor is expected to be 36,000. Fabrication was done with standard computer-numerical-control (CNC) milling techniques, with a nominal positional accuracy of ~10 $\mu$m for each hole.

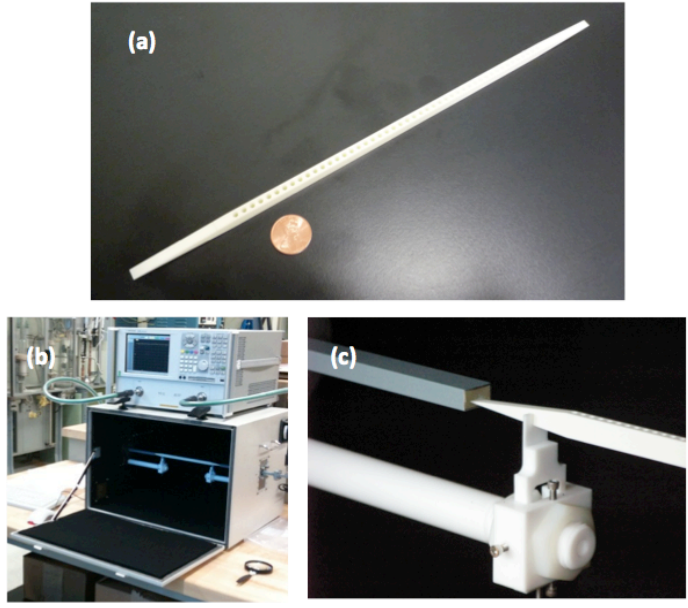

Fig. 2. (a) Images of the fabricated alumina cavity with slant sections at both ends to facilitate coupling to metallic waveguides. (b) Setup for transmittance measurement. (c) Coupling components between the metallic waveguide and the dielectric rod.

The fabricated structure is shown in Fig. 2(a), and the setup used for our two-port transmittance measurements in Fig. 2(b). The device was connected to a network analyzer via conventional WR62 microwave waveguides placed at each end. Because of the geometrical mismatch between the metallic waveguide (15.8mm×7.9mm) and our dielectric waveguide resonator (4.72mm×9.43mm), we designed slant sections at both ends of the resonator to allow partial insertion into the WR62 waveguide and thereby facilitate coupling [Fig. 2(c)]. In Fig. 3(a), we show the amplitude spectrum from the transmission measurement. It demonstrates a large bandgap from 15.4GHz to 17.7GHz, which is consistent with the theory. The transmission coefficient of the cavity mode can be generally fitted using a Fano model [21, 22],

$$t = A\left[\eta\exp(i\varphi) + \frac{1}{1 + i\frac{\omega - \omega_0}{\Delta\omega_{FWHM}/2}}\right] \quad (5)$$

where $A$ is a constant representing system loss, $\eta^2$ denotes the ratio of power coupled to other channels (direct transmission, higher-order modes, etc.), $\varphi$ is the phase difference, $\omega_0$ is the resonant frequency, and $\Delta\omega_{FWHM}$ is the cavity full-width half-maximum linewidth.

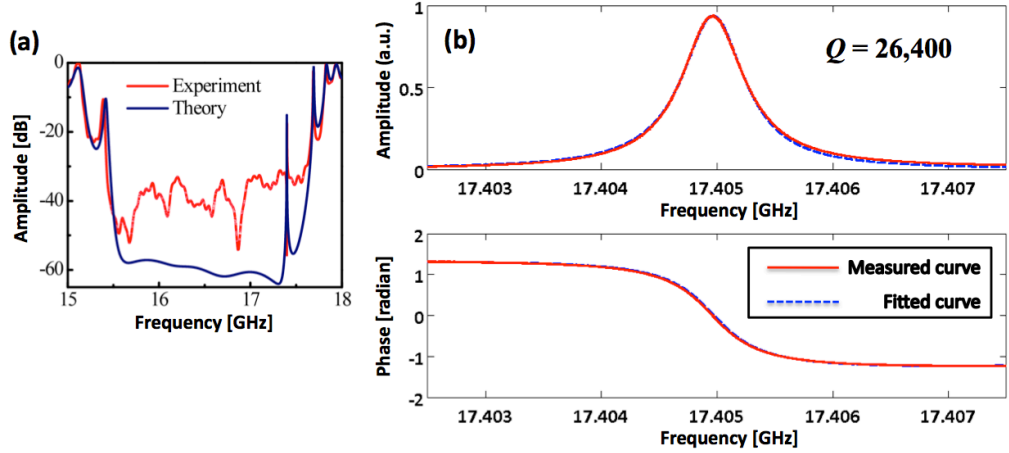

Fig. 3. (a) Amplitude spectrum with a large frequency range from 15GHz to 18GHz, showing the bandgap of the structure. (b) Amplitude and phase spectra of the cavity mode from the transmittance measurement. The dashed curves are fitted with Eq. 5.

Unlike most photonic crystal cavity experiments at optical frequencies, here both amplitude and phase of the transmittance can be obtained. By varying the position of the resonator with respect to the waveguide ports, both $\eta$ and $\varphi$ can be modified, whereas $\omega_0$ and $\Delta\omega_{FWHM}$ stay constant. Here we select a set of data with minimized coupling to direct, non-resonant channels (negligible $\eta$), which results in a Lorentzian resonance lineshape. We plot the resulting spectra in Fig. 3(b), where the cavity resonance is clearly seen at 17.405GHz, in good agreement with the design value. By fitting the measured data with Eq. 5 we obtain a full-width half-maximum linewidth $\Delta\omega_{FWHM}$ of 0.659MHz, corresponding to a $Q$ factor of 26,400, which is slightly smaller than the expected $Q$ value of 36,000. With the resonator end-coupled to a larger WR90 waveguide, we observed $Q$ values up to 29,500 albeit with a less symmetrical line-shape. The observed discrepancy of $Q$ factor can be attributed to higher dielectric losses in the bulk material at this frequency, to fabrication tolerances on the holes position and size and to resonant losses at the transition between waveguide and dielectric. In general, a higher $Q$ factor can be obtained with less lossy materials or at cryogenic temperatures.

Next, we insert a small antenna inside the cavity in order to study the interaction between the cavity mode and a radiating dipole. The antenna is constructed by stripping both the outer conductor and the dielectric core of a coaxial cable (Megaphase Corp., model ClearPath-A06), exposing the inner conductor for a total length of 9 mm. The relatively small diameter (0.305 mm) of the inner conductor was chosen to minimize any perturbation of the cavity mode.

In Fig. 4(a), we plot the reflectance spectrum measured with a network analyzer connected to the above mentioned center-fed antenna, when this is placed in free space (red) and in the center hole of cavity (blue). A clearly visible reflectance dip with a 2.2dB depth at ~17.43GHz is observed, which we interpret as a convincing signature of the Purcell effect of the cavity. When the antenna is coupled to the cavity mode, the radiation rate is enhanced, and thus the load impedance is modified. This results in a variation of the reflectance at the analyzer port, due to the impedance mismatch between the antenna load and the cable. From Fig. 4(a), we also note that there is a change in the resonant frequency of the system, from 17.40GHz to 17.43GHz. This is due to the small but non-negligible perturbation of the metallic antenna, which blue-shifts the resonance. The free-space background in Fig. 4(a) (red-solid curve) arises from the small reflectance between the network analyzer/cable connection, and the cable/antenna connection. This background prevents us from accurately quantifying the load impedance variation and subsequently extracting a numerical value for the Purcell enhancement factor.

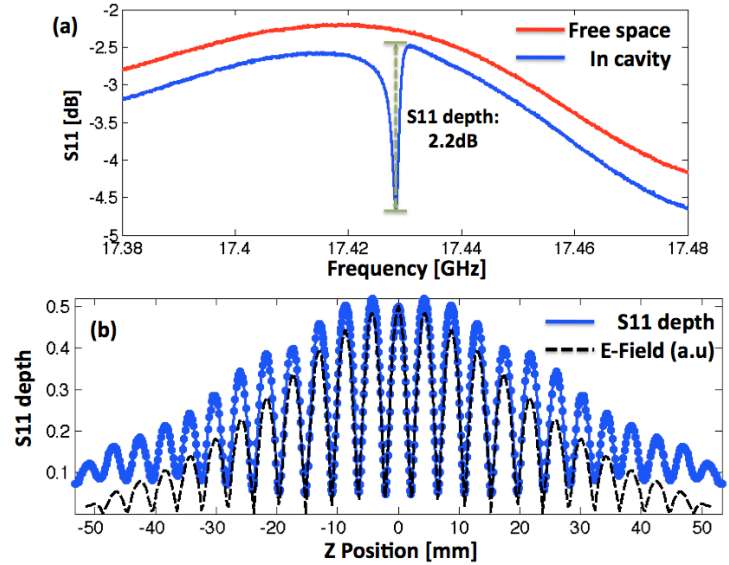

Fig. 4. (a) Reflectance measurement of the center-fed antenna, with and without the cavity. (b) Reflectance depth as a function of the z-position. Also shown (dashed line), the simulated electric-field amplitude along the resonator.

Next, we scanned the antenna in z-direction along the length of the dielectric rod to probe the electric field distribution of the cavity mode. This is analogous to near-field scanning optical microscopy (NSOM) [23, 24]. In Fig. 4(b), we plot the measured reflectance depth as a function of the z-position (blue-solid). In comparison, we also plot the electric field amplitude (black-dash). It can be seen that the two curves follow the same z-dependence, which verifies that the S11 dip we observe is due to coupling to the cavity mode.

## 4. Summary

A high-$Q$ microwave resonator with a mode volume smaller than one cubic wavelength has been designed and fabricated. A record high Quality factor ($Q = 26,400$) for photonic crystal cavities at microwave frequencies has been measured at room temperature. In addition to its ultra-high $Q/V$ factor, the cavity is uniquely designed to have its electric field concentrated in air. Coupling to this cavity mode is enhanced by the Purcell factor. We believe that this device is well suited for conducting microwave cavity QED experiments and for developing a variety of novel microwave devices.

**Acknowledgment**

Y. Zhang would like to dedicate this paper to Mughees Khan and his wife. The authors would also like to acknowledge the generous support from Schlumberger-Doll Research Center at Cambridge, Massachusetts.

**Figure captions**

Fig. 1. (a) Schematic of the device. (b) Diagram of tapered photonic bandgap. (c) Transverse mode profile of the fundamental TM-polarized mode. (d) Mode profile of the cavity mode.

Fig. 2. (a) Images of the fabricated alumina cavity with slant sections at both ends to facilitate coupling to metallic waveguides. (b) Setup for transmittance measurement. (c) Coupling components between the metallic waveguide and the dielectric rod.

Fig. 3. (a) Amplitude spectrum with a large frequency range from 15GHz to 18GHz, showing the bandgap of the structure. (b) Amplitude and phase spectra of the cavity mode from the transmittance measurement. The dashed curves are fitted with Eq. 5.

Fig. 4. (a) Reflectance measurement of the center-fed antenna, with and without the cavity. (b) Reflectance depth as a function of the z-position. Also shown (dashed line), the simulated electric-field amplitude along the resonator.

**References**

1. T. Yoshie, A. Scherer, J. Hendrickson, G. Khitrova, H. M. Gibbs, G. Rupper, C. Ell, O. B. Shchekin, and D. G. Deppe, "Vacuum Rabi splitting with a single quantum dot in a photonic crystal nanocavity," Nature **432**, 200-203 (2004).

2. E. Yablonovitch, T. J. Gmitter, R. D. Meade, A. M. Rappe, K. D. Brommer, and J. D. Joannopoulos, "Donor and Acceptor Modes in Photonic Band-Structure," Phys. Rev. Lett. **67**, 3380-3383 (1991).

3. B. Temelkuran, E. Ozbay, J. P. Kavanaugh, G. Tuttle, and K. M. Ho, "Resonant cavity enhanced detectors embedded in photonic crystals," Appl. Phys. Lett. **72**, 2376-2378 (1998).

4. M. A. Shapiro, W. J. Brown, I. Mastovsky, J. R. Sirigiri, and R. J. Temkin, "17 GHz photonic band gap cavity with improved input coupling," Phys. Rev. Spec. Top-Ac. **4**, 042201 (2001).

5. E. I. Smirnova, A. S. Kesar, I. Mastovsky, M. A. Shapiro, and R. J. Temkin, "Demonstration of a 17-GHz, high-gradient accelerator with a photonic-band-gap structure," Phys. Rev. Lett. **95**, 074801 (2005).

6. E. Di Gennaro, S. Savo, A. Andreone, V. Galdi, G. Castaldi, V. Pierro, and M. R. Masullo, "Mode confinement in photonic quasicrystal point-defect cavities for particle accelerators," Appl. Phys. Lett. **93**, 164102 (2008).

7. C. Jin, B. Cheng, B. Man, Z. Li, and D. Zhang, "Two-dimensional dodecagonal and decagonal quasiperiodic photonic crystals in the microwave region," Phys. Rev. B **61**, 10762-10767 (2000).

8. J. M. Raimond, M. Brune, and S. Haroche, "Colloquium: Manipulating quantum entanglement with atoms and photons in a cavity," Rev. Mod. Phys. **73**, 565-582 (2001).

9. A. Wallraff, D. I. Schuster, A. Blais, L. Frunzio, R. S. Huang, J. Majer, S. Kumar, S. M. Girvin, and R. J. Schoelkopf, "Strong coupling of a single photon to a superconducting qubit using circuit quantum electrodynamics," Nature **431**, 162-167 (2004).

10. A. Petosa, A. Ittipiboon, Y. M. M. Antar, D. Roscoe, and M. Cuhaci, "Recent advances in dielectric-resonator antenna technology," IEEE Antennas Propag. **40**, 35-48 (1998).

11. M. J. R. Previte, and C. D. Geddes, "Fluorescence microscopy in a microwave cavity," Opt. Express **15**, 11640-11649 (2007).

12. Y. Zhang, I. Bulu, T. Botto, B. Levitt, J. Shah, W.-M. Tam, and M. Loncar, "TM-Polarized Air-Band-Mode Photonic Crystal Cavities at Microwave Frequency for cQED Applications," in *International Conference on Photonic and Electromagnetic Crystal Structure* (Granada, 2010).

13. Y. Zhang, M. W. McCutcheon, I. B. Burgess, and M. Loncar, "Ultra-high-Q TE/TM dual-polarized photonic crystal nanocavities," Opt. Lett. **34**, 2694-2696 (2009).

14. J. S. Foresi, P. R. Villeneuve, J. Ferrera, E. R. Thoen, G. Steinmeyer, S. Fan, J. D. Joannopoulos, L. C. Kimerling, H. I. Smith, and E. P. Ippen, "Photonic-bandgap microcavities in optical waveguides," Nature **390**, 143-145 (1997).

15. A. R. M. Zain, N. P. Johnson, M. Sorel, and R. M. De la Rue, "Ultra high quality factor one dimensional photonic crystal/photonic wire micro-cavities in silicon-on-insulator (SOI)," Opt. Express **16**, 12084-12089 (2008).

16. P. B. Deotare, M. W. McCutcheon, I. W. Frank, M. Khan, and M. Loncar, "High

quality factor photonic crystal nanobeam cavities," Appl. Phys. Lett. **94**, 121106 (2009).

17.	Y. Zhang, M. Khan, Y. Huang, J. Ryou, P. Deotare, R. Dupuis, and M. Loncar, "Photonic crystal nanobeam lasers," Appl. Phys. Lett. **97**, 051104 (2010).

18.	Y. Zhang, and M. Loncar, "Ultra-high quality factor optical resonators based on semiconductor nanowires," Opt. Express **16**, 17400-17409 (2008).

19.	P. Lalanne, and J. P. Hugonin, "Bloch-wave engineering for high-Q, small-V microcavities," IEEE. J. Quantum. Elect. **39**, 1430-1438 (2003).

20.	J. D. Joannopoulos, and J. N. Winn, *Photonic crystals: molding the flow of light* (Princeton University Press, 2008).

21.	M. Galli, S. L. Portalupi, M. Belotti, L. C. Andreani, L. O'Faolain, and T. F. Krauss, "Light scattering and Fano resonances in high-Q photonic crystal nanocavities," Appl. Phys. Lett. **94**, 071101 (2009).

22.	Lj. Babic, and M. J. A. de Dood, "Interpretation of Fano lineshape reversal in the reflectivity of photonic crystal slabs, Opt. Express **18**, 26569-26582 (2010).

23.	U. Durig, D. W. Pohl, and F. Rohner, "Near-Field Optical-Scanning Microscopy," J. Appl. Phys. **59**, 3318-3327 (1986).

24.	L. Lalouat, B. Cluzel, P. Velha, E. Picard, D. Peyrade, J. P. Hugonin, P. Lalanne, E. Hadji, and F. de Fornel, "Near-field interactions between a subwavelength tip and a small-volume photonic-crystal nanocavity," Phys. Rev. B **76**, 041102 (2007).